\documentclass[review]{elsarticle}
\usepackage{lineno}
\usepackage{amsmath}
\usepackage{longtable}
\usepackage{epstopdf}

\journal{.....}

\begin{document}  

\title{Unusual charge exchange by swift heavy ions at solid surfaces}
\author[]{Tapan Nandi \corref{cor1}}
\author[]{Prashant Sharma \corref{cor2}}
\ead{Present address: Faculty of Physics, Weizmann Institute ofScience, Rehovot 7610001, Israel, phyprashant@gmail.com}
\author[]{Pravin Kumar}
\address{Inter University Accelerator Centre, Aruna Asaf Ali Marg, New Delhi - 110067, INDIA}
\begin{abstract}

We have employed x-ray spectroscopy to probe  the charge changing process  only in the bulk of the foil when swift heavy ions pass through it. In contrast, the electromagnetic methods take into account integral  effect of the charge changing process in the bulk as well as the charge exchange phenomenon at the surface of the foil. Thus, the difference between  the mean charge states so measured from the two methods disentangles the charge exchange phenomenon at the surface from the charge changing process in the bulk and, provides opportunities to refine the understanding of ion-surface interactions. Very surprisingly, up to tens of electrons per event participate in the charge exchange phenomenon during swift heavy ion-surface interactions. This finding has been validated with a series of experiments using several ions (z = 22-35) in the energy range of 1.5-3.0 MeV/u and also verified theoretically with Fermi-gas model. Interestingly, such unusual charge exchange phenomenon could play significant  role in  x-ray emission of many astrophysical environments, infrared emission bands from range of environments in galaxies, accelerator physics, ion energy losses in solids, heavy ion cancer treatments, inner shell ionization by heavy ions, and  surface modifications in nano scale. 
\end{abstract}

\maketitle
 \indent The charge exchange (CX) phenomenon, i.e., non-radiative electron capture  (NRC) and radiative electron capture (REC), is very important aspect of ion-atom (gas) as well as ion-solid collisions. The number of electrons captured by the projectiles in ion-atom collisions is dictated  by its charge state as well as target nuclear charge. However, the capture process in ion-solid collisions is independent of target nuclear charge owing to dangling electrons at the solid surface. Irrespective of initial charge state, slow highly charged ions ($v_p<v_0, v_p$=ion velocity and $v_0$=Bohr velocity) tend to acquire neutrality after passage through a thin foil  [1]. Furthermore, the relative contribution of the electron capture process in the bulk and at foil surface remains unaccounted. In order to investigate only the surface contribution, the scattering of multi charged slow ions incident on a metal surface at glancing angles is considered [2]. Even though,  highly charged relativistic ions at normal incidence are found to capture several electrons after passage through a thin foil [3], the net electron capture contribution at the exit surface remains still unknown. To deduce it experimentally, a clear distinction between charge states of the incident and emerging ions is necessary and hence, multi-electron heavy ions are preferred as projectiles. This, in turn, enables us to distinguish ion-surface interactions at the entrance and exit surfaces.  In this letter, a clear evidence of the net electron capture contribution at the exit surface is showcased with the multi-electron, swift heavy ions (SHI).\\
\indent At intermediate energies ($v_p>v_0$), the x-ray spectra of highly charged ions in post foil interactions account for charge states much higher [4,5] than that measured usually by electromagnetic methods [6]. The two methods are distinguishable in terms of the distance between source and detector. This distance for beam-foil spectroscopy is of the order of mm, whereas the detector is placed a few meters away from the source for electromagnetic methods. To analyze the charge state of the multi-electron projectile ions right at the beam-foil interaction region, the x-ray spectroscopy method [7,8] is utilized. Several electrons from the entrance surface  are captured by the projectile ions in their high Rydberg states, which are re-ionized along with ionization of the projectile ions at high frequency, ion-solid collisions in the bulk. 
The higher charge state of the emerging ions so formed undergo multi-electron capture (MEC) processes in the high Rydberg states at the exit surface of the foil [9]. The high Rydberg states are long lived and therefore, the signature of their existence is hardly traceable in x-ray spectra recorded for ion-solid collision region. The REC process does not undergo any excitation mechanism and thus, is instantaneous, producing no lifetime structure. Also, REC 
structure is distinguishable from primary x-ray peaks.  Therefore, the ions undergone to REC process further experience the interaction with the exit surface to capture the electrons in high Rydberg states. Hence, the primary peak in the x-ray spectra recorded during the ion-solid collisions demonstrate the charge changing processes only in the bulk. 
Whereas, the electromagnetic methods [6] measure integral  effect of the charge changing process in the bulk as well as the charge exchange phenomenon at the surface of the foil. Therefore, the difference between the mean charge state measured only in  bulk ($q^b_m$) and, in both the bulk and exit surface ($q^t_m$), can shed light on the CX at the exit surface. 
\\
\indent Interestingly, an improved empirical formula constructed from a large set of experimental charge-state distributions analyzed by electromagnetic methods [10]   shows good agreement with experimentally measured integral effect in charge change of the projectile ions. To the best of our knowledge, no theory has been developed until now to quantify the charge changing process in the bulk owing to the scarcity of the measurements [11]. Recently, an extensively used x-ray spectroscopy technique [8], which rescues all challenges, has been employed to determine the $q^b_m$ for several ions ($z$ = 22-35) in the energy range of 1.5-3.0 MeV/u. With great surprise, we notice that simple Fermi-gas model explains entire data set very well.  This development enables us to assess theoretically the CX contribution only at the exit surface.
\\
\indent Several ion-solid collision experiments were carried out using 15 UD tandem accelerator at IUAC. The energetic ion beams of $^{48}$Ti, $^{51}$V, $^{56}$Fe, $^{58}$Ni, $^{63}$Cu,  and $^{79}$Br (current $\approx$ 1 to 2 pnA) were allowed to pass through a carbon foil to attain charge state distribution (CSD) at equilibrium thickness. To avoid influence of target modifications, the foils with similar thickness were used in each experimental run. In the present work, we have used the characteristic K-$\alpha$ x-ray emissions to determine the q$^b_m$ of the projectile ions. For this purpose, widely used experimental technique [7,12,13] is employed and its details is given elsewhere [8]. During experiments, the pressure  in the vacuum chamber was maintained around 1 $\times$ 10$^{−6}$ Torr. Two germanium ultra-low-energy detectors (GUL0055 and GUL0035, Canberra Inc., with a 25 μm thick Be entrance window, resolution 150 eV at 5.9 keV, and constant detector efficiency in the range of 5 to 20 keV) were placed at right angle to the beam axis. While, the target foil holder was kept at 45$^{\circ}$ to the beam axis and a set of collimators were used to restrict the scattered x-rays recorded by the detector. Hence, only the prompt x rays are allowed to enter into the detectors through thin mylar windows of 6 $\mu$m thickness. The Doppler broadening is maximum for this geometry, whereas the first-order Doppler shift is zero
and, depending on the beam energies, the second-order shift appears at the fourth or fifth decimal place confirming the requirement of no further corrections. The detectors were calibrated with different radioactive sources like $^{55}$Fe, $^{57}$Co, and $^{241}$Am before as well as after the experiments to check the systematic errors in the spectra and no significant deviations were noticed. The NRC processes  at the exit of the target surface involve high Rydberg states and do not influence the prompt x-ray spectra at all. Thereby, in contrast to the conventional electromagnetic methods accounted for  aggregated  effects of charge changing processes in the bulk as well as the charge  exchange phenomenon in the exit surface, the present x-ray technique showcases an account for the charge changing processes only in the bulk. 

\begin{figure}[!t]
\centering
\includegraphics[scale=0.25]{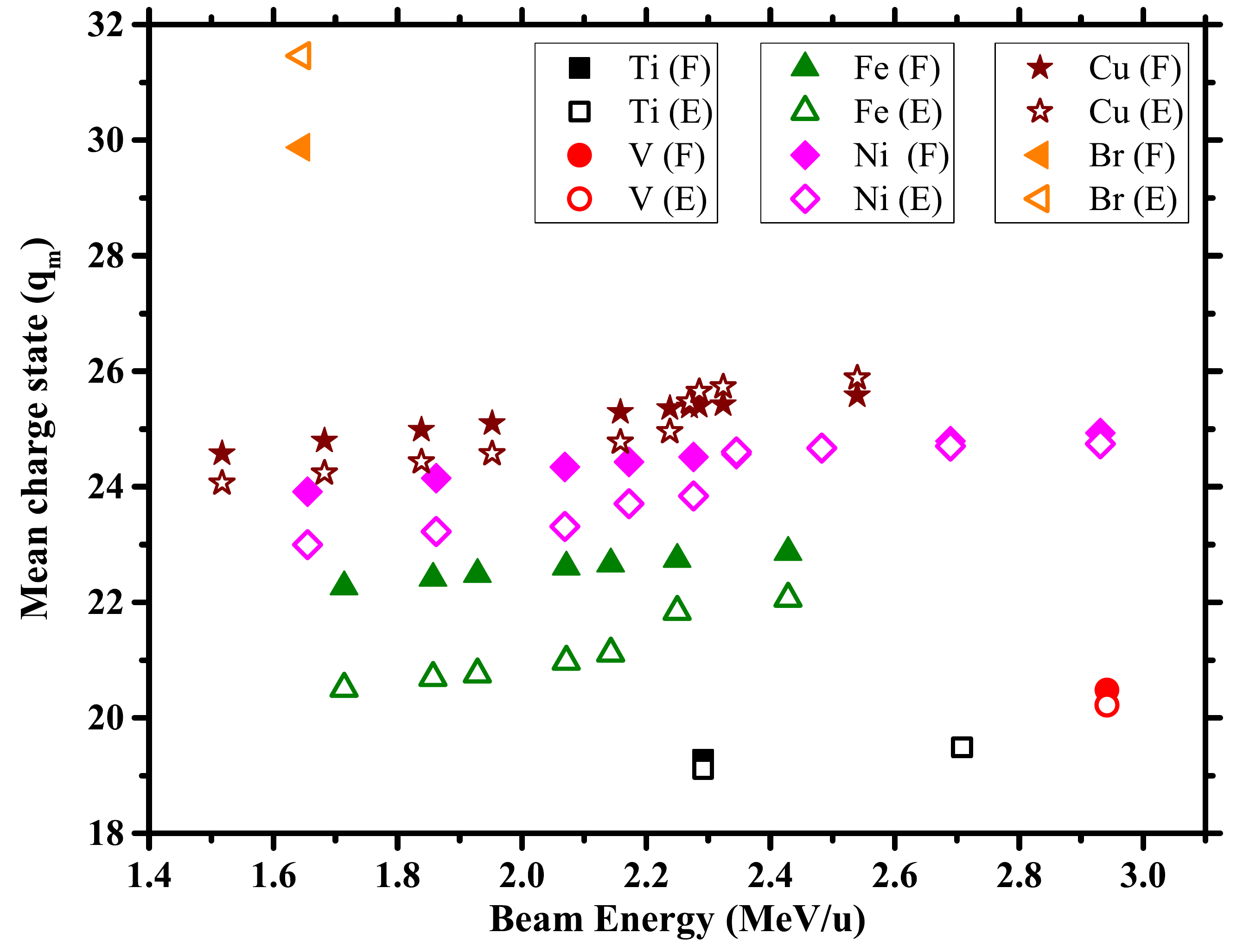}
\caption{\label{fig1f} Comparison of experimentally governed (empty symbols) and theoretically estimated (filled symbols) $q^b_m$s for various projectile ions as a function of beam energies. The Fermi-gas model was used for theoretical estimations.  The projectile ions are specified with their elemental symbols. E and F in the parentheses represent the experiment and Fermi-gas model, respectively. The uncertainties are  within the symbol size.} 
\end{figure}
\indent A series of ion-solid collision experiments were carried out with various ion species and taking the advantage of analysis made in our earlier studies [8,13,14], the mean charge state of the projectile ions in the bulk of the carbon foil was measured from the centroid of the K$\alpha$ x-ray peaks. The fitting error in the centroid energy is found to be less
than 1\%. The amorphous carbon foils used in all experiments showed excellent electrical conductivity and could be approximated to the metal behaving like an electron gas at high temperature and pressure. In these beam-foil experiments, the swift ions interact with bulk Fermi electrons (velocity $\approx v_F$) [15]. Further, the impact of interaction depends upon the beam velocity ($v_p$) as well and the $q^b_{m}$, which is related with two velocities as:
\begin{equation} \label{Fermi-gas}
q^b_{m} = z_p(1-\dfrac{v_F}{v_p}).
\end{equation}
\begin{figure}[!t]
\centering
\includegraphics[scale=0.25]{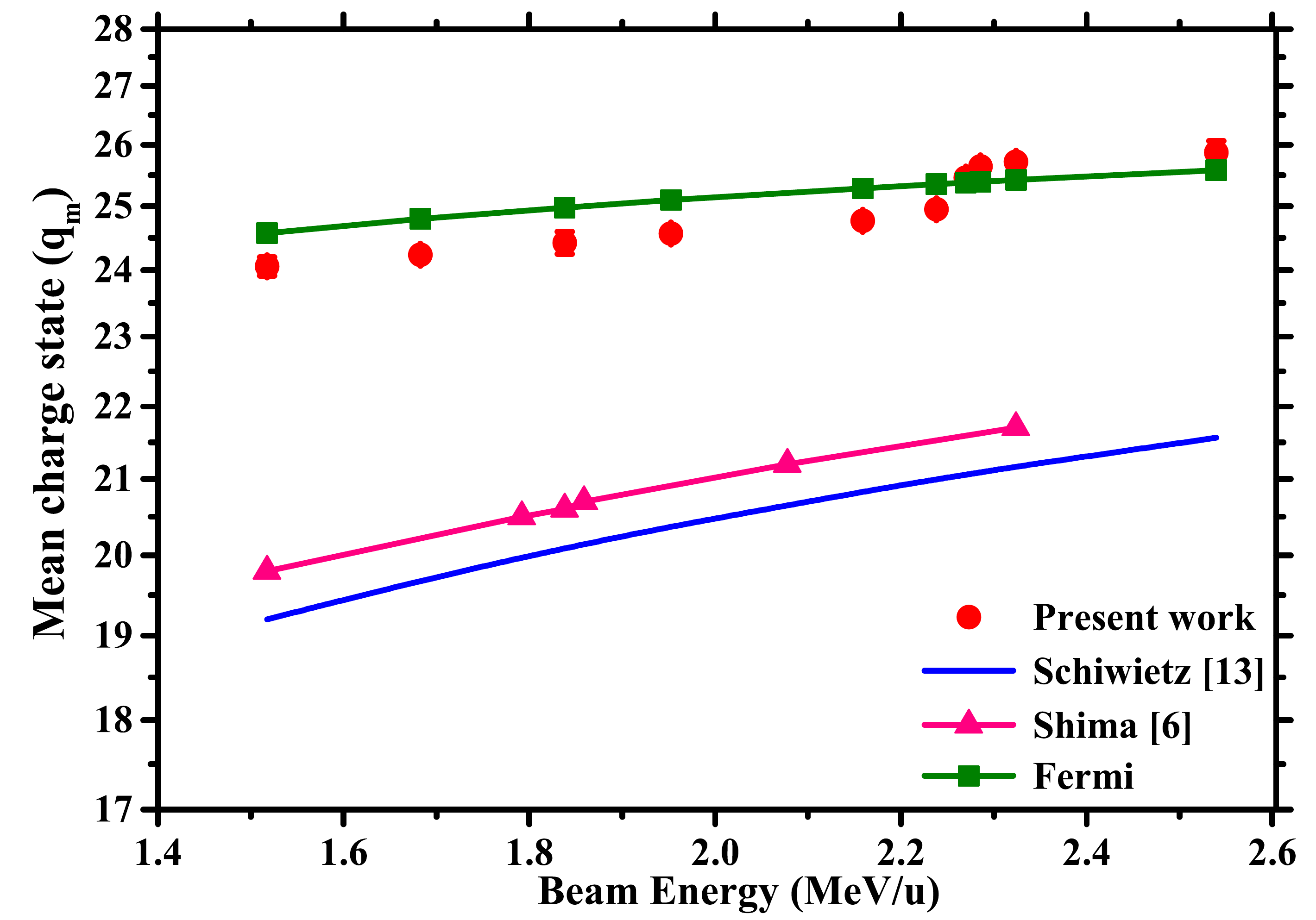}
\caption{\label{fig1f} 
Comparison of the $q^b_{m}$ and $q^t_{m}$ as measured by x-ray spectroscopy (present work)  and electromagnetic  techniques, respectively. The Cu beam of various energies was allowed to pass through a carbon foil. The uncertainties are  within the symbol size.}
\end{figure}
\begin{figure*}[!t]
\centering
\includegraphics[scale=0.25]{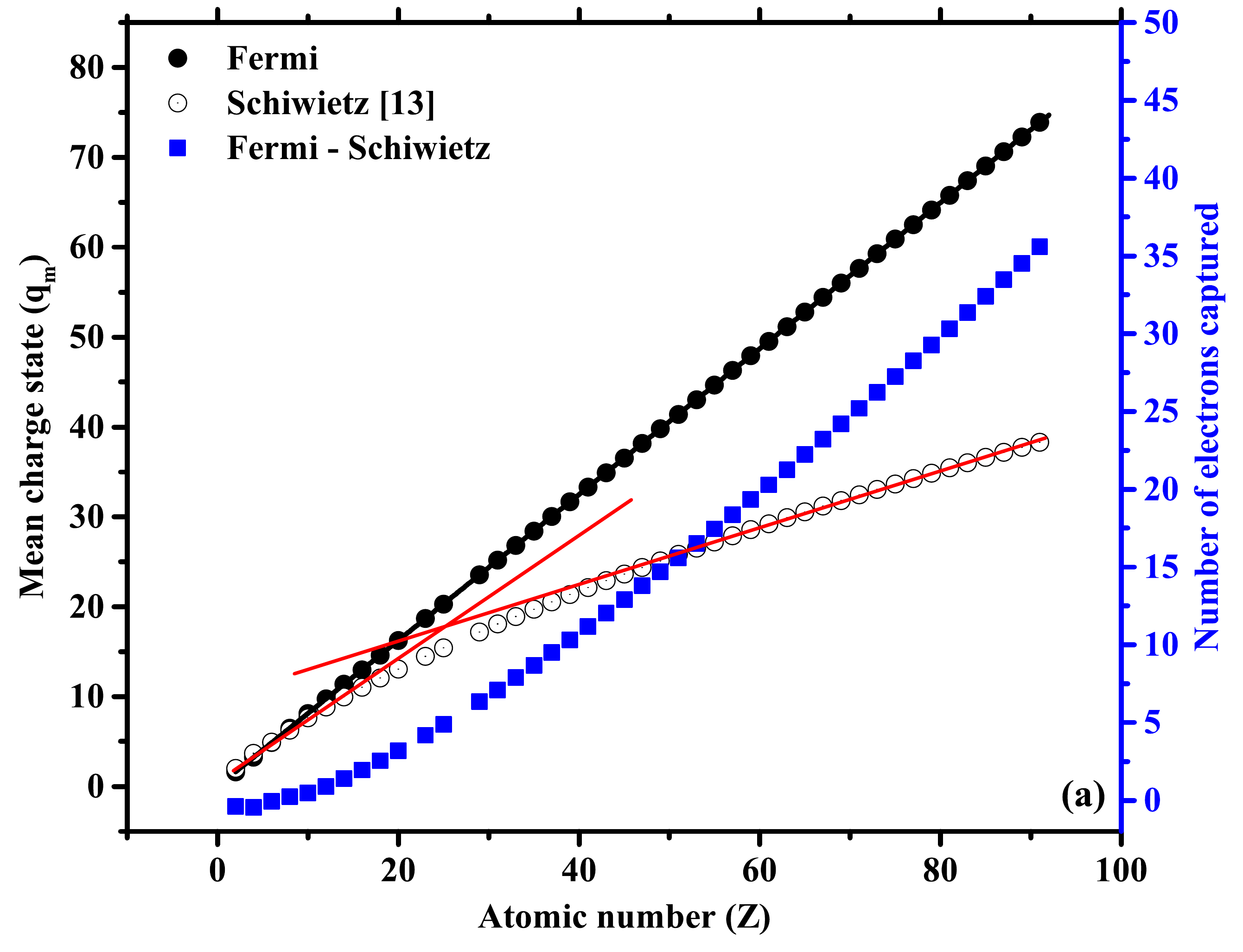} \includegraphics[scale=0.25]{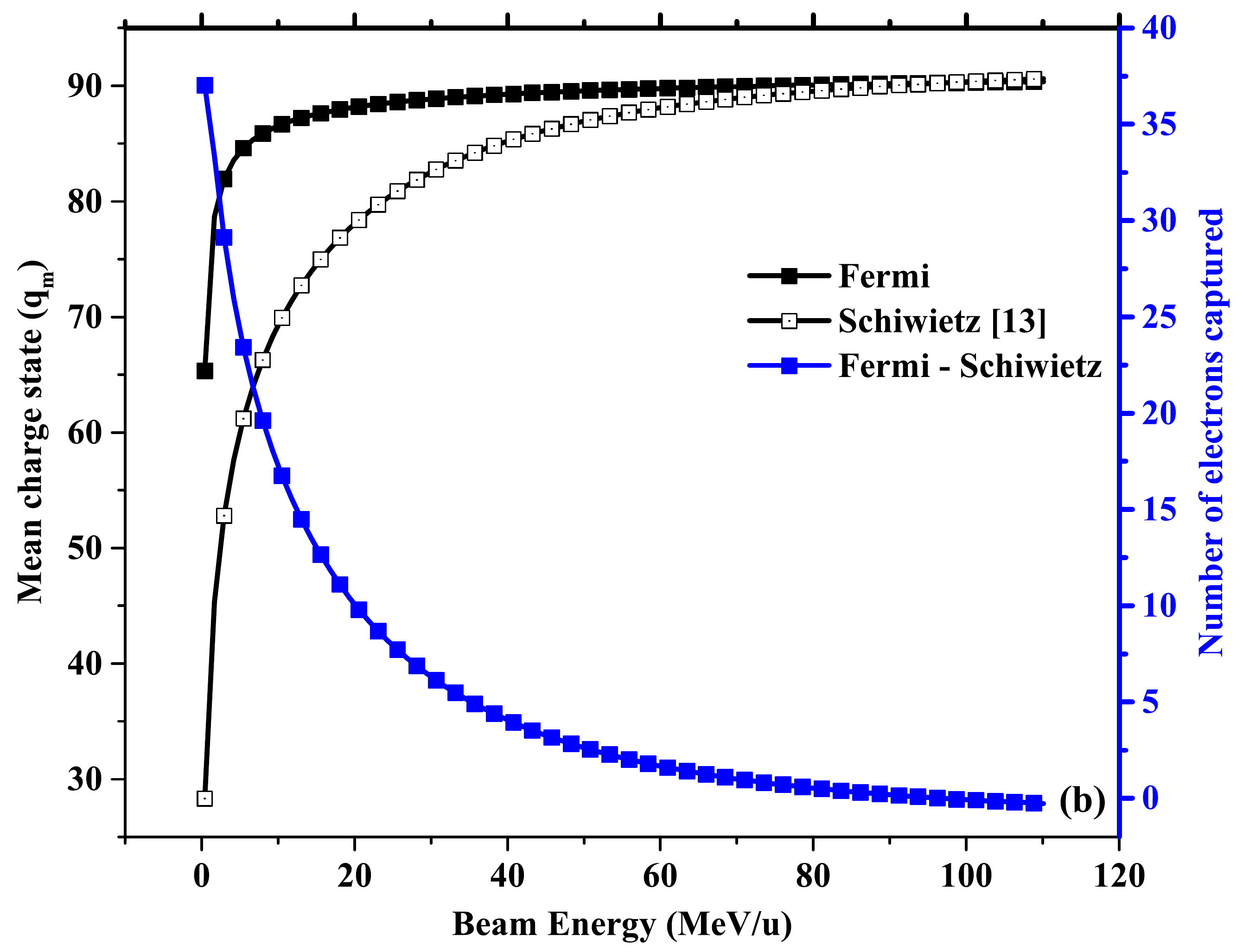}
\caption{\label{fig1f} Theoretical values of $q^b_{m}$ from Fermi-gas model and $q^t_{m}$ from improved Schiwietz  formula as a function of (a) atomic number up to uranium with constant beam energy of 1 MeV/u and (b) beam energies (1-110 MeV/u) for $^{235}U$ ions. The difference between the $q^b_{m}$ and $q^t_{m}$ representing a direct measure of the charge exchange at the exit surface of the foil ($N^e_c$) is also shown.  The analysis shows that as high as 37 electrons are captured from the exit surface by 1 MeV/u $^{235}$U ion and this number decreases with energy; reduces to nearly zero for energies $>$ 80 MeV/u. }
\end{figure*}
 \begin{figure*}[!h]
\centering
\includegraphics[scale=0.061]{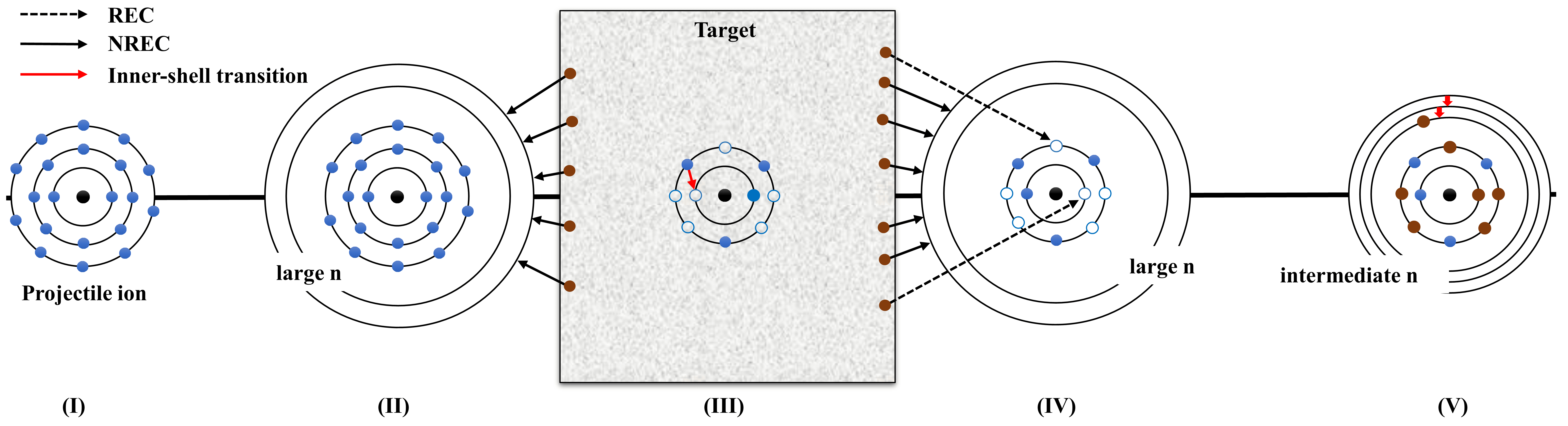}
\caption{\label{1schm} The illustration of a five-stage model (see text) based on the
charge changing process in the bulk and charge exchange phenomenon  at foil surfaces during the passage of projectile ions.}
\end{figure*}
Where $z_P$ is the nuclear charge of the projectile ions and $v_F$=2.61$\times 10^6$ m/sec for amorphous carbon foil [16]. The theoretical values of $q^b_{m}$ using equation (1) along with experimental values are depicted in Fig. 1.
The theoretical predictions match well with the experiments and, agreement improves further for beam energies $>$ 2 MeV/u. The dynamic screening of the projectile charge by the electrons in the solid target [15] and transient electric field acting on the heavy ions during deceleration in the dense medium [17] seem to put a small perturbation on projectile charge evolution inside the solids. Equation (1) is valid only for $v_p>v_F$ and equilibrium foil thickness, for which a number of collisions take place and the projectile charge attains a kind of saturation.

\indent We also made an attempt to compare the $q^b_{m}$ with $q^t_{m}$. As  $q^t_{m}$ can be either measured by electromagnetic methods or estimated accurately using an improved formula [10], a case study of the copper beam on account of mean charge state using various approaches is summarized in Fig. 2. The measured $q^b_{m}$ values are in well accord with the predictions of Fermi-gas model and $q^t_{m}$ values [6] follow Schiwietz formula [10]. The difference between $q^b_{m}$ and $q^t_{m}$ gives a direct measure of the charge exchange at the exit surface of the foil ($N^e_c$) ignoring the effect of autoionization at the moment. \\
\indent The energy loss by SHI is mostly in the bulk and only a small fraction of it takes place at the exit surface [18]. The loss in both regions depends on the charge state of the ion in the bulk [19,20], which is far different from its value during incidence as well as emergence of ions. This behaviour  is well explained by the Fermi-gas model and hence finds wide applications in important research fields, for example, calculations of stopping power  [19,21] and innershell ionization in solids [22]. Also, this information is useful in calculation of the biologically effective dose not only in the target region but also in the entire irradiation volume during heavy ion cancer treatment [23]. \\
\indent In Fig.3 (a) and (b), we have plotted $q^b_{m}$, $q^t_{m}$ and $N^e_c$ as a function of $z_p$ and beam energy (keeping one constant at a time), respectively. It can be seen that $N^e_c$ takes a significant figure for $z_p$ $\approx 15$ and increases with increase in $z_p$ up to the value of uranium. The analysis shows that up to 37 electrons are captured at the exit surface of the foil for uranium ions at 1 MeV/u and this number decreases rapidly with increasing the beam energy owing to reduced NRC cross-section. For example, $N^e_c$ is only a few for 46.5 MeV/u, which is in good agreement with the experiment [3]. Large number of electrons can be captured to the projectile ion if enough vacancies are available. This condition prevails in the circular Rydberg states and after large-scale electron capture, the hallow ions are formed. In comparison to such hallow ion formation, the capture of two or more electrons in low lying states is more complex because of Coulomb screening and electron-electron correlations.   \\
\indent Because of the fact that Auger rates are high for low $z_p$, low for high $z_p$, and moderate for intermediate $z_p$, the data points governed by Schiwietz formula are fitted well with two straight lines as shown in Fig.3(a).  With similar analogy, $N^e_c$ reflects the variation of Auger and radiation decay rates with $z_P$. In contrast, the $q^b_{m}$ data fit well with a straight line implying that the charge changing process in the bulk is independent of the decay modes because fast Auger transition is not possible at high frequency, ion-solid collisions. \\ 
\indent In astronomical environments, the CX is an important aspect of atomic collisions in gas phase [24]. Besides, it plays a significant role in ion-solid collisions because of the existence of amorphous and crystalline silicates, refractory oxides of magnesium, aluminum, silicon, and iron in the majority of O-rich envelopes, and, SiC in C-rich and AGB stars [25]. Further, amorphous organic solids, widely observed in a range of environments in the galaxies [26] are also heavily influenced by CX. Apart from this, CX is significant in basic and applied physics, viz., surface modification of solids by SHI at nano scales  [27-30]. \\
\indent As bulk and surface for gas targets are undefined, the concept of only mean charge state (q$_{m}$), which is analogous to $q^t_m$ in solids, is introduced. Generally, the $q^t_m$ in solid is expected to be higher than that in gas target [10] because of the high density. However, this fact is not true as in experiments, rather a bit higher value of $q_m$ is observed  [31]. This effect is attributed to the CX at exit surface in solids. Therefore, the surface contribution is extremely important in deciding the charge strippers for accelerators [32]. \\
\indent Having charge exchange at the exit surface, the resulting ions have to elapse a long while before they are counted in the detector placed at the focal plane. In this duration, the $z_p$ dependent auto-ionizing decay is highly probable. As the difference of $q^b_{m}$ and $q^t_{m}$ is also $z_p$ dependent, the net number of electrons transferred (net charge-exchange) as measured by the detector is governed by the charge exchange as well as autoionization processes. The former lowers the projectile charge state and the latter increases the same. Therefore, actual charge exchange at the exit surface of the foil can be larger than the net charge exchange, $N^e_c$ = $q^b_{m}$ - $q^t_{m}$, depending upon $z_p$. For instance, Auger rates are so large for the light ions that all the excited states can be autoionized before they are detected. Hence,  $N^e_c\approx0$  till $z_p \approx15$ and it increases with $z_p$ owing to increasing radiative rates. The $N^e_c$ for high $z_p$, say, for Uranium, stands for the charge exchange at the exit surface of the foil as the Auger rate is almost zero.\\
\indent The charge changing process in the bulk material and charge exchange phenomenon at the exit surface can be summarized in a five-stage model as shown in Fig. 4. In the first stage, the ions have to spend a long while before approaching the target and hence, lie in the ground state. In the second stage, before entering to the target, the entrance surface electrons are polarized and pulled out by the projectile ions and finally, are placed into Rydberg states. In the third stage, such states will be re-ionized in the high-frequency collisions inside the bulk. Here, many more electrons lying in the ground state are also ionized enhancing the projectile charge state. Further, innershell vacancies are created in the high-frequency collisions and thus, the observation of K-x rays. In the fourth stage, as highly charged ions cross the exit surface, many electrons can be picked up and placed in the high lying states including the Rydberg and circular Rydberg states. The properties of the states so formed can differ depending on $z_p$. Up to $z_p$ = 15, Auger rates are many orders of magnitude higher than radiative decay rates.  For $z_p$ $\approx$ 25, the  Auger or radiative decay rates are comparable. In contrast, for $z_p$  $\ge$ 47, radiative decay rates are many orders of magnitude higher than Auger rates and the high lying states so formed are termed as circular Rydberg states, possessing highest magnetic quantum number as observed in several experiments [9,33]. These states undergo neither Auger transition nor fast radiative decay because of restriction on the  dipole selection rules $\Delta l= \pm 1$. They decay only through a cascading chain and take a considerable time to relax into the ground states. The decay through chain also continues during flight between source and detector. If time taken by the projectile ions to reach to the detector is more than the cascading time, i.e., $T(n, l)>\sum \tau(n,l)$, where $\tau(n,l)$ is the lifetime of the circular Rydberg state with quantum numbers $n$ and $l$, the ground state is achieved. Otherwise, the state lies somewhere in the middle of the cascading chain as shown in the fifth stage of the model. Whatsoever, the ionic state of projectile remains intact as seen by the electromagnetic analyzer. \\
\indent To conclude, the electromagnetic methods for charge state analysis provide an integral measure of the charge changing processes in the bulk and charge exchange phenomenon at the exit surface of the foil. However, disentangling these two contributions are essential for many applications, e.g., x-ray emission of many astrophysical objects, the infrared emission bands from range of environments in the galaxies, accelerator physics, ion energy losses in solids, cancer therapy and ionization by heavy ions, and the surface modifications in nano scales. Accordingly, we have employed the x-ray spectroscopy technique to measure the mean charge states of swift heavy ions evolved due to the charge changing process only in the bulk of the carbon foils. We find that the mean charge states so measured by the two methods are very different, because x-ray technique takes account of $q^b_{m}$, whereas $q^t_{m}$ is deduced by the electromagnetic analyzer;  the $q^b_{m}$ being higher than $q^t_{m}$. Theoretical prediction of $q^b_{m}$ are made using a simple model assuming that the target electrons form a Fermi-gas, with which the swift heavy ions interact. For a series of measurements with several ions (z = 22-35) in the energy range 1.5-3.0 MeV/u, a very good agreement is seen between the present experiments and theory. The $q^t_{m}$s of the ions as measured by electromagnetic methods are also evaluated accurately by an improved formula. The $q^b_{m}$ - $q^t_{m}$ is a measure of the net charge exchange process responsible at the exit surface of the foil. Very surprisingly, for 1 MeV/u uranium ions, up to tens of electrons per event participate in charge exchange process at the surface of the carbon foils.  
 
\section*{Acknowledgement}
 We would like to acknowledge the co-operation and support received from the Pelletron accelerator staff and all colleagues of the atomic physics group, IUAC, New Delhi.

\section*{Bibliography}
[1] M. Hattass, T. Schenkel, A. Hamza, A. Barnes, M. Newman, J. McDonald,
T. Niedermayr, G. Machicoane, D. Schneider, Charge equilibration time of slow, highly charged ions in solids, Physical review letters 82 (24) (1999) 4795.\newline
[2] H. Winter, Image charge acceleration of multi-charged argon ions in grazing collisions with an aluminum surface, EPL (Europhysics Letters) 18 (3) (1992) 207.\newline
[3] H. Brauning, P. Mokler, D. Liesen, F. Bosch, B. Franzke, A. Kramer,
C. Kozhuharov, T. Ludziejewski, X. Ma, F. Nolden, M. Steck, T. Stohlker,
R. Dunford, E. Kanter, G. Bednarz, a. Warczak, Z. Stachura, L. Tribedi,
T. Kambara, D. Dauvergne, R. Kirsch, C. Cohen, Physical Review Letters
86 (6) (2001) 991{994.\newline
[4] T. Nandi, P. Marketos, P. Joshi, R. Singh, C. Safvan, P. Verma, A. Mandal,
A. Roy, R. Bhowmik, Physical Review A 66 (5) (2002) 052510.\newline
[5] T. Nandi, N. Ahmad, A. Wani, P. Marketos, Reliable measurement of the
li-like ti 22 48 1 s 2 s 2 p p 5/ 2 o 4 level lifetime by beam-foil and beam{
two-foil experiments, Physical Review A 73 (3) (2006) 032509.\newline
[6] K. Shima, T. Ishihara, T. Miyoshi, T. Mikumo, Equilibrium charge-state
distributions of 35146-mev cu ions behind carbon foils, Physical Review A
28 (4) (1983) 2162.\newline
[7] J. P. Santos, A. M. Costa, J. P. Marques, M. C. Martins, P. Indelicato, F. Parente, Physical Review A 82 (6) (2010) 062516.\newline
[8] P. Sharma, T. Nandi, X-ray spectroscopy: An experimental technique to
measure charge state distribution during ionsolid interaction, Physics Letters A 380 (12) (2016) 182--187.\newline
[9] T. Nandi, Formation of the Circular Rydberg States in Ion-Solid Collisions,
The Astrophysical Journal 673 (2008) L103--L106.\newline
[10] G. Schiwietz, P. L. Grande, Nuclear Instruments and Methods in Physics
Research B 175177 (0) (2001) 125--131.\newline
[11] A. Lifschitz, N. Arista, Effective charge and the mean charge of swift ions
in solids, Physical Review A 69 (1) (2004) 012902.\newline
[12] J. T. Schmelz, R. Brown, The sun: a laboratory for astrophysics, Vol. 373, Springer Science \& Business Media, 2012.\newline
[13] P. Sharma, T. Nandi, Shakeoff ionization near the coulomb barrier energy,
Physical review letters 119 (20) (2017) 203401.\newline
[14] P. Sharma, T. Nandi, Experimental evidence of beam-foil plasma creation
during ion-solid interaction, Physics of Plasmas 23 (2016) 083102.\newline
[15] W. Brandt, R. Laubert, M. Mourino, A. Schwarzschild, Dynamic screening
of projectile charges in solids measured by target x-ray emission, Physical
Review Letters 30 (9) (1973) 358.\newline
[16] M.-Z. Huang, W. Ching, Electronic and transport properties of perfect sp
2-bonded amorphous graphitic carbon, Physical Review B 49 (7) (1994)
4987.\newline
[17] J. Lindhard, A. Winther, Nuclear Physics A 166 (3) (1971) 413--435.\newline
[18] T. Nandi, K. Haris, G. Singh, P. Kumar, R. Kumar, S. Saini, S. Khan,
A. Jhingan, P. Verma, a. Tauheed, D. Mehta, H. Berry, Physical Review
Letters 110 (16) (2013) 163203.\newline
[19] C. Frey, G. Dollinger, A. Bergmaier, T. Faestermann, P. Maier-Komor,
Charge state dependence of the stopping power of 1 mev/a 58Ni ions, Nuclear Instruments and Methods in Physics Research B 107 (1-4) (1996) 31--35.\newline
[20] F. Gruner, F. Bell, W. Assmann, M. Schubert, Integrated approach to the
electronic interaction of swift heavy ions with solids and gases, Physical
review letters 93 (21) (2004) 213201.\newline
[21] P. Sigmund, Charge-dependent electronic stopping of swift nonrelativistic
heavy ions, Physical Review A 56 (5) (1997) 3781.\newline
[22] D. Burch, N. Stolterfoht, D. Schneider, H. Wieman, J. Risley, Projectile
charge-state dependence of ne k-shell ionization and fluorescence yield in
50-mev cl n++ ne collisions, Physical Review Letters 32 (21) (1974) 1151.\newline
[23] M. Kramer, M. Scholz, Treatment planning for heavy-ion radiotherapy: calculation and optimization of biologically effective dose, Physics in Medicine \& Biology 45 (11) (2000) 3319.\newline
[24] S. Kwok, Nature 430 (7003) (2004) 985.\newline
[25] L. M. Ziurys, The chemistry in circumstellar envelopes of evolved stars:
Following the origin of the elements to the origin of life, Proceedings of the
National Academy of Sciences 103 (33) (2006) 12274--12279.\newline
[26] S. Kwok, Y. Zhang, Mixed aromatic{aliphatic organic nanoparticles as carriers of unidentied infrared emission features, Nature 479 (7371) (2011) 80.\newline
[27] Z. Wang, C. Dufour, E. Paumier, M. Toulemonde, The sensitivity of metals under swift-heavy-ion irradiation: a transient thermal process, Journal
of Physics: Condensed Matter 6 (34) (1994) 6733.\newline
[28] F. Aumayr, S. Facsko, A. S. El-Said, C. Trautmann, M. Schleberger, Single ion induced surface nanostructures: a comparison between slow highly
charged and swift heavy ions, Journal of Physics: Condensed Matter 23 (39)
(2011) 393001.\newline
[29] Z. Siwy, E. Heins, C. C. Harrell, P. Kohli, C. R. Martin, Journal of the American
Chemical Society 126 (35) (2004) 10850--10851.\newline
[30] Y. Cheng, H. Yao, J. Duan, L. Xu, P. Zhai, S. Lyu, Y. Chen, K. Maaz,
D. Mo, Y. Sun, et al., Nanomaterials 7 (5) (2017) 108.\newline
[31] R. Clark, I. Grant, R. King, D. Eastham, T. Joy, Equilibrium charge state
distributions of high energy heavy ions, Nuclear Instruments and Methods
133 (1) (1976) 17--24.\newline
[32] P. Scharrer, C. E. Dullmann, W. Barth, J. Khuyagbaatar, A. Yakushev,
M. Bevcic, P. Gerhard, L. Groening, K. Horn, E. Jager, et al., Measurements of charge state distributions of 0.74 and 1.4 Mev/u heavy ions passing through dilute gases, Physical Review Accelerators and Beams 20 (4)(2017) 043503.\newline
[33] A. P. Mishra, T. Nandi, B. Jagatap, Resonances in the population of circular rydberg states formed in beam-foil excitation, Journal of Quantitative
Spectroscopy and Radiative Transfer 120 (2013) 114--119.\newline

\end{document}